\documentclass{GR24}
\usepackage{lmodern}

\usepackage{amsmath,amssymb,amsfonts,amsthm}
\usepackage{physics}
\usepackage{xcolor}
\usepackage{graphicx}
\usepackage{hyperref}
\usepackage{cleveref}

\usepackage{tikz}
\usetikzlibrary{decorations.pathmorphing,patterns}

\usepackage{subcaption}
\usepackage{wrapfig}

\newcommand{\td}{\text{d}}

\newcommand{\im}{{\rm i}}

\begin{document}

\title{Singularity Resolution of Quantum Black Holes in (A)dS}

\author{Sofie Ried$^{1}$}

\affil{$^1$School of Mathematical and Physical Sciences, University of Sheffield, Sheffield, United Kingdom}

\email{sried1@sheffield.ac.uk}

\begin{abstract}
The singularities present at the centre of black holes signal a break down of the classical theory. In this paper, we demonstrate a resolution of the Schwarzschild-(Anti-)de Sitter singularity by imposing unitary evolution with respect to unimodular time. Employing the Henneaux-Teitelboim formulation of unimodular gravity, we perform a canonical quantization on a symmetry-reduced Schwarzschild-(Anti-) de Sitter model. This leads to a Wheeler-DeWitt equation that effectively becomes a Schrödinger equation in unimodular time. By imposing unitarity, we discover a family of quantum theories in which the classical singularity is resolved. These theories each allow only semiclassical states corresponding to one mass sign: either positive, negative, or zero. Furthermore, we derive an analytical expression for the quantum-corrected Schwarzschild metric, which is modified by a new length scale $r_{min}$ that governs the black hole's transition to a white hole.
\end{abstract}

\section{Introduction}
Singularities in black holes mark a breakdown of classical general relativity, where geodesics of infalling objects reach an end and predictability is lost. The semiclassical effect of Hawking radiation implies black holes evaporate, making the singularity's physics relevant to outside observers and leading to the information loss paradox \cite{Hawking:1975vcx}.

Including quantum gravitational effects might resolve singularity issues by ensuring unitary time evolution, preventing wave functions from ending abruptly. In this framework, trajectories could extend beyond classical singularities. However, general relativity's covariance means coordinate time lacks physical significance. Relational dynamics uses a degree of freedom as a clock for unitary evolution, maintaining gauge invariance but introducing clock dependency in the quantum theory \cite{gotay_quantum_1983,gotay_remarks_1997}.

In this conference paper we will summarise the methods and findings of \cite{Gielen:2025ovv}, as well as have a closer look at the resulting quantum-corrected Schwarzschild metric studied in more detail in \cite{Gielen:2025}. We will introduce unimodular time and motivate it as a good choice of clock variable. We will see how imposing unitarity with respect to unimodular time resolves the black hole singularity in the setting of a Schwarzschild symmetry reduced minisuperspace model.

\section{Theoretical framework}
\label{sec:theory}

\subsection{Unimodular gravity}
\label{sec:unig}
We will be working in the Henneaux-Teitelboim formulation of unimodular gravity \cite{henneaux_cosmological_1989}, which is classically equivalent to general relativity. By promoting the cosmological constant $\Lambda$ to a field and then adding a Lagrange multiplier term to the Einstein-Hilbert action we arrive at:
\begin{align}
\label{equ:HT}
    S_{HT}[g_{\mu\nu},\Lambda,T^{\mu}] = \frac{1}{2}\int\td^4x[\sqrt{-g}R-2\Lambda(\sqrt{-g}-\partial_{\mu}T^{\mu})]\,.
\end{align}
This introduces four new degrees of freedom in terms of $T^{\mu}$, since we are only considering the total divergence three of them are gauge. Variations with respect to $T^{\mu}$ and $\Lambda$ result in the following equations of motion:
\begin{align}
   \partial_{\mu}\Lambda=0\,, \quad\sqrt{-g} = \partial_{\mu}T^{\mu}\,.
\end{align}
The first one ensures that we get the usual cosmological constant as an integration constant and the second one relates the new degree of freedom to the determinant of the metric. The latter is called the \textit{unimodular condition}. Another thing to note from \cref{equ:HT}, is that $\Lambda$ is the conjugate momentum of $T^{0}$. This will become important when quantising, since $\Lambda$ will correspond to $-\im\hbar\partial_{T^0}$ and $T^0$ is directly related to our clock, unimodular time. 

Unimodular time is defined as the integral of $n_{\mu}T^{\mu}$ over a hypersurface $\Sigma(t)$, where $n_{\mu}$ is the future directed normal to that hypersurface:
\begin{align}
    T_{\Lambda}(t) := \int_{\Sigma(t)}\td^3x\; n_{\mu}T^{\mu}\,.
\end{align}
It's interpretation as time stems from the fact that the difference of unimodular time of two hypersurfaces is given by the 4-volume between them. This can easily be seen from the definition of unimodular time and using the unimodular condition and Stokes theorem. In the next section we will make certain symmetry assumptions, in particular homogeneity, such that $ T:=T_{\Lambda}=VT^{0}$ where $V$ is the volume of the spatial hypersurface (which we will have to compactify to make this well defined).

\subsection{Schwarzschild symmetry reduced metric}
\label{sec:mini}
Since we are interested in quantising the Schwarzschild spacetime with a cosmological constant, we will make use of its symmetries to drastically reduce the degrees of freedom before quantising. The classical metric we are interested in has the following form:
\begin{align}
\label{equ:lineel}
    \td s^2 = -\frac{\eta(t)N(t)^2}{\xi(t)}\td t^2 + \frac{\xi(t)}{\eta(t)} \td z^2 + \eta(t)^2\td\Omega^2\,.
\end{align}
Where $\td\Omega^2$ is the standard metric on $\mathbb{S}^2$ and $N$ is the lapse function. Here $t$ is the time coordinate inside the horizon ($\xi>0$) and $z$ is the time coordinate in the exterior ($\xi<0$). Unlike for $\xi$ we will fix the sign of $\eta\geq 0$, as it can be interpreted as the radius of the two-spheres.

\section{Canonical quantisation}
\label{sec:can}
Plugging the symmetry reduced metric (\ref{equ:lineel}) into the action (\ref{equ:HT}) we can calculate the Hamiltonian:
\begin{align}
\label{equ:ham}
    H = -N\left[\frac{1}{\eta^2}\pi_{\eta}\pi_{\xi}+\frac{1}{\eta^2}-\Lambda\right] = 0\,,
\end{align}
which is constraint to vanish, as usual in diffeomorphism invariant settings. By canonically quantising this constraint we arrive at the Wheeler--DeWitt equation
\begin{align}
\label{equ:WdW}
    \im\hbar\partial_T\psi = \hat{H}_S\psi\quad \text{with}\quad\hat{H}_S = \frac{\hbar^2}{\eta^2}\partial_{\eta}\partial_{\xi}-\frac{1}{\eta^2}\,.
\end{align}
Where we have defined the Schrödinger Hamiltonian $\hat{H}_S$ in such a way, that the Wheeler--DeWitt equation takes the form of a Schrödinger equation in unimodular time $T$.

We now compute expectation values of operators like $\hat{\eta}$ and $\hat{\xi}$ at a fixed unimodular time $T$. Since $T$ is a spacetime field, not a coordinate, expressions like $\eta(T)$ are gauge invariant. Requiring unitary evolution in $T$ amounts to making $\hat{H}_S$ self-adjoint. Remarkably, its expectation value equals $-\Lambda$, so this also promotes $\hat{\Lambda}$ to a quantum observable.

\section{Results}
\label{sec:res}

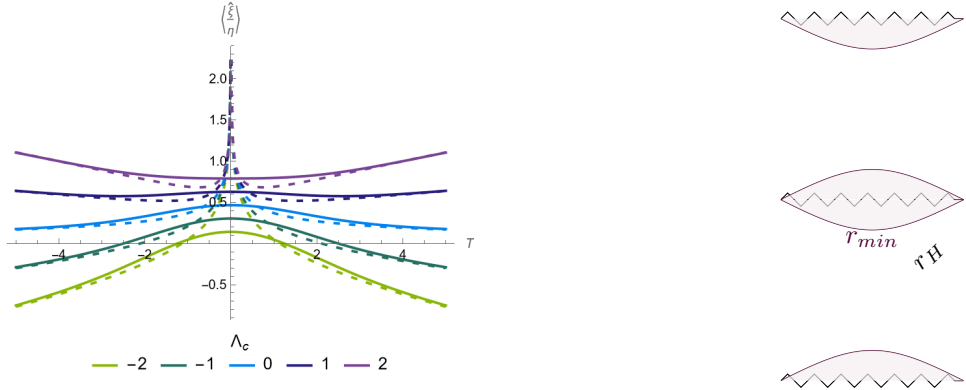
\begin{figure}
    \centering
\begin{subfigure}[b]{0.48\textwidth}
    \centering
    \includegraphics[width=0.8\linewidth]{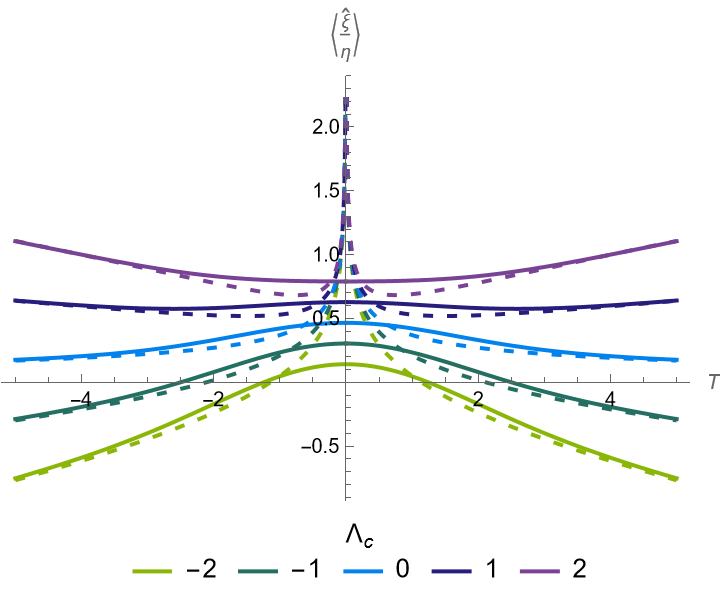}
    \caption{Solid lines show $\langle\widehat{\xi/\eta}\rangle<\infty$ everywhere. Dashed lines show the classical black hole ($T<0$) and white hole ($T>0$) solutions.}
    \label{fig:evsf}
\end{subfigure}
\hfill
\begin{subfigure}[b]{0.48\textwidth}
    \centering
\begin{tikzpicture}[scale=0.4]
    \colorlet{mydarkpurple}{blue!40!red!50!black}
    \colorlet{mylightpurple}{mydarkpurple!80!red!6}

    \coordinate (Il) at (0,-3);
    \coordinate (Ib) at (3,-6);
    \coordinate (Ir) at (6,-3);
    \coordinate (It) at (3,0);
    \coordinate (IIt) at (-3,0);
    \coordinate (IIl) at (-6,-3);
    \coordinate (IIb) at (-3,-6);
    \coordinate (Vr) at (6,3);
    \coordinate (Vt) at (3,6);
    \coordinate (Vl) at (0,3);
    \coordinate (VIt) at (-3,6);
    \coordinate (VIl) at (-6,3);

    \draw -- (Il) -- (Ib) --  (Ir) --  (It) -- node[midway, below, sloped] {$r_H$} (Il) -- cycle;

    \draw -- (IIl) -- (IIb) -- (Il) -- (IIt) -- (IIl) -- cycle;

    \draw -- (Vl) -- (It) -- (Vr) -- (Vt) -- (Vl) -- cycle;

    \draw -- (VIl) -- (IIt) -- (Vl) -- (VIt) -- (VIl) -- cycle;



    \draw[decorate,decoration=zigzag] (IIb) -- (Ib);

    \draw[decorate,decoration=zigzag] (IIt) -- (It);

    \draw[decorate,decoration=zigzag] (VIt) -- (Vt);

    \tikzset{declare function={
        kruskal(\x,\c)  = {\fpeval{asin( \c*sin(2*\x) )*2/pi}};
    }}

    \draw[mydarkpurple,line width=0.4,samples=20,smooth,variable=\x,domain=-3:3] 
        plot(\x,{-3*kruskal((\x+3)*pi/12,0.5)})
        plot(\x,{ 3*kruskal((\x+3)*pi/12,0.5)});

    \fill[fill=mylightpurple, opacity=0.7]
        plot[smooth,domain=-3:3,variable=\x] 
        (\x,{-3*kruskal((\x+3)*pi/12,0.5)})
    --
        plot[smooth,domain=-3:3,variable=\x] 
        (\x,{ 3*kruskal((\x+3)*pi/12,0.5)})
    -- cycle;

    \draw[mydarkpurple,line width=0.4,samples=20,smooth,variable=\x,domain=-3:3] 
        plot(\x,{3*kruskal((\x+3)*pi/12,0.5)-6});

    \fill[fill=mylightpurple, opacity=0.7]
        plot[smooth,domain=-3:3,variable=\x] 
        (\x,{3*kruskal((\x+3)*pi/12,0.5)-6})
    -- (Ib) -- (IIb)
    -- cycle;

     \draw[mydarkpurple,line width=0.4,samples=20,smooth,variable=\x,domain=-3:3] 
        plot(\x,{-3*kruskal((\x+3)*pi/12,0.5)+6});

    \fill[fill=mylightpurple, opacity=0.7]
        plot[smooth,domain=-3:3,variable=\x] 
        (\x,{-3*kruskal((\x+3)*pi/12,0.5)+6})
    -- (Vt) -- (VIt)
    -- cycle;

    \node[mydarkpurple] at (0,-1.3) {$r_{min}$};
\end{tikzpicture}
\caption{Causal structure of the quantum corrected Schwarz\-schild spacetime. The purple region is "cut out".}
\label{fig:conformaldiagram}
\end{subfigure}
\caption{Figure (\ref{fig:evsf}) shows the resolution of the singularity in the quantum theory and \cref{fig:conformaldiagram} illustrates the spacetime structure described by the quantum corrected Schwarzschild metric.}
\end{figure}

We will now present the main results of the analysis of the quantum system using the following semiclassical states (with normalisation factor $\mathcal{N}$):
\begin{align}
\label{equ:alphasc}
    \psi( \eta,\xi,T) &= \int_{-\infty}^{\infty}\frac{\td\Lambda}{2\pi}\int_{-\infty}^{\infty}\frac{\td k}{2\pi}\,\alpha_{sc}(\Lambda,k)\,e^{\im\Lambda T}e^{\im\left[k\xi+\frac{1}{k}\left(\frac{\Lambda}{3}\eta^3-\eta\right)\right]}\,,\\
    \text{with}\quad\alpha_{sc}(\Lambda,k) &= \mathcal{N} e^{-\im\frac{\beta}{2} k} e^{-\frac{(|k|-k_c)^2}{2\sigma_k^2}}e^{-\frac{(\Lambda-\Lambda_c)^2}{2\sigma_{\Lambda}^2}}\,,
\end{align}
characterised by the expectation values $k_c$ and $\Lambda_c$ of $\hat{\pi}_{\xi}$ and $\hat{\pi}_T = \hat{\Lambda}$ and their variance. The parameter $\beta$ is the self-adjoint extension parameter. It arises when looking for self-adjoint extensions of $\hat{H}_S$. There is no unique extension, instead the theory depends on a free function $\chi(k)$. Using symmetry arguments we can narrow it down to a one parameter family of quantum theories, indexed by $\beta$.

\subsection{Singularity resolution}
\label{sec:singres}
No matter the choice of $\beta$, the singularity is resolved in all quantum theories. This can be seen in \cref{fig:evsf} as the expectation value of the scale factor stays finite where the classical solution diverges. Singularity resolution is a generic feature of unitary theories in which the clock reaches in the singularity in finite time, as was first discussed in \cite{gotay_quantum_1983,gotay_remarks_1997}.

\subsection{Mass sign of semi-classical states}
\label{sec:masssign}
Our second key result is the link between the semiclassical black hole mass and the self-adjoint extension parameter $\beta$, for which $\frac{1}{2}\beta k_c^2 = 2GM$ holds. For fixed $\beta$, the mass magnitude can vary with $k_c$, but its sign is fixed: $\text{sgn}(M) = \text{sgn}(\beta)$. This ensures all semiclassical states in the theory share the same mass sign, avoiding issues like vacuum instability otherwise present in singularity free theories \cite{HorowitzMyers}.  

\subsection{Quantum-corrected Schwarzschild metric}
\label{sec:metric}
By replacing $\eta$ and $\xi$ in \cref{equ:lineel} with the corresponding expectation values we can write down a semiclassical quantum corrected Schwarzschild-(Anti-)de Sitter metric:
\begin{align}
    \td s^2=-\frac{\td T^2}{\langle\eta(T)\rangle^3 \langle\xi(T)\rangle}+\frac{\langle\xi(T)\rangle}{\langle\eta(T)\rangle}\td z^2+\langle\eta(T)\rangle^2 \td\Omega^2\,.
\end{align}
Because the full expressions for the expectation values are quite lengthy and in the spirit of semiclassicality we will further assume $\sigma_k\ll k_c$. In addition we will restrict ourselves to the asymptotically flat case $\Lambda_c = 0$. After performing a coordinate transformation to the usual Schwarzschild coordinates we arrive at the following metric:
\begin{align}
    {\rm d}s^2 &= - \left(1-\frac{r_H}{r}\right){\rm d}z^2 +\frac{{\rm d}r^2k_c^2\,T'(r)^2}{r^4\left(1-\frac{r_H}{r}\right)} + r^2\,{\rm d}\Omega^2\,,\\
    \frac{k_c}{r^2}T'(r)&\approx1-\frac{1}{6}\left(\frac{\sqrt{\pi}r_{min}}{\Gamma\left(\frac{2}{3}\right)r}\right)^{6}-\frac{1}{2}\left(\frac{\sqrt{\pi}r_{min}}{\Gamma\left(\frac{2}{3}\right)r}\right)^{12}\,.
\end{align}
Where $r_H = 2GM$ is the usual Schwarzschild event horizon. We can see that the $g_{zz}$ component of the metric is exactly the same as in the classical theory, but the $g_{rr}$ component gets corrections from $\frac{k_c}{r^2}T'(r)$. The function $T(r)$ is only implicitly defined as the inverse function of a confluent hypergeometric function, but can be expanded for large $r$ as written above. The correction terms are governed by a new length scale $r_{min}$ given as a function of the state defining parameters in the following way:
\begin{align}
    r_{{\rm min}}= \frac{\Gamma\left(\frac{2}{3}\right)}{\sqrt{\pi}}\left(\frac{3k_c}{\sigma_\Lambda}\right)^{1/3} \,.
\end{align}
This minimum radius is the scale at which the black hole transitions to a white hole geometry. For more details on the quantum corrected Schwarzschild metric see \cite{Gielen:2025}.

\section{Summary and conclusion}
\label{sec:con}
Imposing unitarity in unimodular time resolves the singularity in a symmetry-reduced Schwarzschild model. All semiclassical states share the same mass sign, and we obtain analytic expressions for expectation values, yielding a Schwarzschild metric with explicit quantum corrections. Deviations from the classical metric are controlled by a minimal length scale $r_{min}$; see \cite{Gielen:2025ovv,Gielen:2025} for details.

\section{Acknowledgements}
\label{sec:ack}
I appreciate the opportunity to speak at GR24-Amaldi16 and am grateful to everyone who attended and contributed to the discussion.

\bibliography{GR24}

\end{document}